\documentstyle[12pt]{article}
\begin{document}
\openup6pt

\title {General relativity limit of the scalar-tensor theories for traceless matter field}
\author{A. Bhadra\thanks{Email address: aru\_bhadra@yahoo.com} \\
High Energy and Cosmic Ray Research Centre\\
University of North Bengal, Siliguri 734430 
INDIA
}
\date{}
\maketitle

\begin{abstract}
$\omega(\phi) \rightarrow \infty$ limit of scalar tensor theories are studied for traceless matter source. It is shown that the limit $\omega(\phi) \rightarrow \infty$ does not reduce a scalar tensor theory to GR. An exact radiation solution of  scalar tensor cosmology under modified Nordtvedt conditions is obtained for flat Friedmann universe.       
\end{abstract}

\section*{I. Introduction}
   Recently Damour and Nordtvedt [1] has been demonstrated quite generally that scalar tensor (ST) theories generically contain a natural attractor mechanism toward general relativity (GR). Such a possibility was also suggested previously [2,3] but only for some particular classes of models. This means ST theories are cosmologically evolved toward a state with no scalar admixture to gravity during matter dominated era [1] or even during radiation era [3,4,5]. This attractor feature is important because it diminishes the doubt on the existence of massless scalar field with gravitational-strength coupling by providing reasonable explanation for the small effect of the scalar fields (in comparison with curvature effect) in the present epoch without requiring any fine-tuned large value of the coupling parameter as needed in the case of celebrated Brans-Dicke (BD) theory [6]. It is worth mentioning that a scalar partner to gravity is inevitable in most efforts at unifying gravity with the other interactions, such as superstring theories [7] or Kaluza-Klien theories [8]. In cosmology also, the scalar field is found to play an appreciable role by providing a natural way to overcome the fine tuning problem of "new" inflationary model of universe by terminating the inflationary era through nucleation of bubbles [9]. The present acceleration of the universe, as revealed from the observational data of the type 1a supernova [10], also can be accommodate with the scalar field coupled gravity theories without the need of any new kind of matter field (so called "Quintessence matter") having a positive energy density but a negative pressure [11,12].\\
The attractor mechanism is based on the idea that a ST theory converges to GR for certain values of the characteristic coupling parameter $\omega(\phi)$, which represents the strength of the coupling between the scalar field and the curvature. During cosmological evolution, the Einstein frame coupling parameter $\alpha(\tilde{\phi})$ tends to zero [1] or equivalently in Jordan frame $\omega(\phi) \rightarrow \infty$ ($\alpha^{2}(\tilde{\phi}) \equiv \frac{1}{3+2\omega(\phi)}$) and this limit is usually regarded as GR limit of ST theories. However, recently it has been shown that the BD theory, which is the simplest ST theory with constant $\omega$, does not necessarily reduce to GR in the limit $\omega \rightarrow \infty$ when the trace of the matter energy-momentum tensor vanishes [13-16]. It is, therefore, important to examine the $\omega(\phi) \rightarrow \infty$ limit of ST theories. \\
In the present work, we investigate $\omega(\phi) \rightarrow \infty$ limit of ST theories in the radiation (and vacuum) universe. We shall rigorously show that the limit $\omega(\phi) \rightarrow \infty$ does not reduce a ST theory to GR. The cosmological evolution of the scale factor a(t) during radiation era will also be analyzed, in presence of a scalar field satisfying the modified (explain below) Nordtvedt conditions [17]. \\
The paper is organized as follows: the general field equations of ST theories are set up in the next section. The non-convergence of ST theories to GR is discussed in sec. III. An exact solution of scale factor $a(t)$ in radiation universe is obtained in sec. IV for general ST theories under modified Nordtvedt conditions, and the results are discussed in sec. V.     \\
\section*{II. Field equations}
Here we consider the most general ST theories of gravity with a single scalar field. In the Jordan conformal frame (since experimentally observed quantities are those written in the Jordan frame, we shall work in this frame throughout the paper), the general form of the action describing a massless  scalar-tensor theory of gravity in natural units ($G=c=1$) is [17,18] 
\begin{equation}
{\cal A} = \frac{1}{16 \pi }\int\sqrt{-g} d^{4}x \left[\phi R-\frac{\omega(\phi)}{\phi} g^{\mu\nu} \nabla_{\mu}\phi \nabla_{\nu}\phi +16\pi{\cal L}_{m} \right ]       
\end{equation}
where R is the Ricci scalar constructed from the metric $g_{\mu\nu}$, and $ {\cal L} _{m} $ is the Lagrangian density of ordinary matter which could include electromagnetic field, nuclear field etc. The principle of equivalence is guaranteed by requiring that the matter field Lagrangian can depend only on the metric $g_{\mu\nu}$ but not on $\phi$. A variation of (1) with respect to $ g^ {\mu \nu} $ and $\phi$ yields, respectively, the field equations
\begin{equation} 
R_{\mu\nu} -\frac{1}{2}g_{\mu\nu}R= \frac{8 \pi} {\phi } T_{\mu \nu}
+\frac {\omega}{\phi ^{2}}\left( \phi_{,\mu} \phi_{,\nu}- \frac{1}{2}g_{\mu\nu} 
\phi^{,\sigma} \phi_{,\sigma} \right) + 
\frac{1}{\phi} \left( \phi_{,\mu;\nu}-g_{\mu \nu} \Box \phi \right),  
\end{equation}
\begin{equation}
\Box\phi=\frac{8\pi}{2\omega(\phi)+3}T-\frac{\omega^{\prime}}{2\omega(\phi)+3}g^{\mu\nu} \phi_{,\mu}\phi_{,\nu} \; ;
\end{equation}
with the energy momentum conservation equation
\begin{equation}
T^{\mu\nu}_{;\nu}=0 \;,
\end{equation}
where $\Box \equiv g^{\mu\nu}\nabla_{\mu}\nabla_{\nu}$, $T$=$T_{\mu}^{\mu}$ is the trace of the matter energy momentum tensor and $\omega^{\prime}\equiv \frac{d\omega}{d\phi}$.  It is clear that the scalar field plays the role which in GR played by the gravitational constant, but with $\phi$ now a dynamical variable.\\
To proceed further, we consider a homogenous and isotropic universe. The line element has then a Robertson-Walker form:
\begin{equation}
ds^{2}=-dt^{2}+a^{2}(t)\left[\frac{dr^2}{1-kr^{2}}+r^2d\Omega^{2} \right] \; ;
\end{equation}
curvature constant $k=-1,0,+1 $ for an open, flat or closed universe respectively. The condition of homogeneity implies that the scalar field is only a function of the time coordinate, $\phi \equiv \phi(t)$. For the above line element, the Eq. (4) reduces to 
\begin{equation}
\dot{\rho}+3\frac{\dot{a}}{a}(\rho +p)=0
\end{equation} 
where an over dot denotes $d/dt$.
For a radiation universe with equation of state $P=\frac{1}{3}\rho$, the above equation yields  
\begin{equation}
\rho=\frac{3}{8\pi}\Gamma a^{-4}
\end{equation}
where $\Gamma$ is a positive constant. $\Gamma=0$ corresponds to vacuum ($\rho=P=0$).
The field equations (2) and (3) thus yield 
\begin{equation}
\frac{\dot{a}^{2}}{a^{2}}=\frac{\Gamma}{\phi a^{4}}-\frac{k}{a^{2}}-\frac{\dot{\phi}}{\phi}\frac{\dot{a}}{a}+\frac{\omega(\phi)}{6}\frac{\dot{\phi}^{2}}{\phi^{2}} \;,
\end{equation}
\begin{equation}
2\frac{\ddot{a}}{a}+\frac{\dot{a}^{2}+k}{a^{2}}=-\frac{\Gamma}{\phi a^{4}}-\frac{\omega(\phi)}{2}\frac{\dot{\phi}^{2}}{\phi^{2}}-2\frac{\dot{\phi}}{\phi}\frac{\dot{a}}{a} -\frac{\ddot{\phi}}{\phi}  \; ,
\end{equation}
and
\begin{equation}
\ddot{\phi}+\frac{3\dot{a}}{a}\dot{\phi}=-\frac{\dot{\phi}^{2}\omega^{\prime}(\phi)} {3+2\omega(\phi)} \;.
\end{equation}
When $\phi$ is a constant, the above equations (8) and (9) reduce to the GR equations with a gravitational constant $G=\frac{1}{\phi}$. 
\section*{III. Non-convergence of ST theories to GR}
The first integral of the equation (10) immediately gives,
\begin{equation}
\dot{\phi}=\frac{A}{a^{3}\sqrt{(3+2\omega(\phi))}}
\end{equation}
where A is a constant.
The above equation is true for {\it any} radiation or vacuum cosmological Friedmann solution and is {\it valid for all ST theories}. It is to be noted that since $A$ is independent of time, it can not be a function of $\omega(\phi)$ but for Brans-Dicke theory ($\omega$ is constant) $A$ may depend on $\omega$. Inserting $\dot{\phi}$ from Eq. (11) into Eqs. (8) and (9) we get under the limit $\omega(\phi) \rightarrow \infty$ 
\begin{equation}
\frac{\dot{a}^{2}+k}{a^{2}}=\frac{\Gamma}{\phi a^4}+\frac{A^{2}}{12 \phi^{2} a^{6}} \;,
\end{equation}
and 
\begin{equation}
2\frac{\ddot{a}}{a}+\frac{\dot{a}^{2}+k}{a^{2}}= -\frac{\Gamma}{\phi a^4} -\frac{A^{2}}{4 \phi^{2}a^{6}}(1-\frac{\omega^{\prime}(\phi)}{\omega^{2}(\phi)}) \;.
\end{equation}
and the Ricci scalar in the same limit is given by 
\begin{equation}
R\simeq -\frac{1}{\phi}\left(\frac{3\omega^{\prime}(\phi)}{4\omega^{2}(\phi)} -\frac{1}{2\phi}\right)\frac{A^{2}}{a^{6}} \; ,
\end{equation}
When a scalar-tensor theory converges to GR, the scalar curvature must approach to zero and at the same time, the scalar field must become a constant as well. It is evident from Eq. (11) that as $\omega (\phi) \rightarrow \infty$, $\phi$ is tending to a constant value ($1$). $A$ should not vanish  in this limit since $A$ is independent of $\omega(\phi)$. $a(t)$ must be bounded for any realistic theory of gravity. Then it follows from Eq. (14) that $R$ is not approaching its GR value ($0$) for large value of $\omega(\phi)$ or even for the Nordtvedt conditions [17] {\it viz}. $\omega(\phi) \rightarrow \infty \; , \frac{\omega^{\prime}(\phi)}{\omega^{3}(\phi)}\rightarrow 0 $. These conditions are essential if the solar system tests are to accord with observations. In fact the first term at the right hand side of Eq. (14) is ambiguous under Nordtvedt conditions. On the other hand, the second term at the right hand side of Eq. (14) becomes non-zero constant as $\omega (\phi)\rightarrow \infty$. 
The said ambiguity in the expression of $R$ under original Nordtvedt conditions can be avoided by slightly modifying the Nordtvedt conditions as 
\begin{equation}
\omega (\phi)\rightarrow \infty \; , \frac{\omega^{\prime}(\phi)}{\omega^{2}(\phi)}\rightarrow 0 
\end{equation}
But even for such modified conditions, $R$ remains non-zero and the field equations are clearly different from the corresponding GR equations with the same energy momentum tensor. {\it This shows that GR cannot be recovered from ST theories by imposing constraints on the coupling parameter $\omega(\phi)$}. \\
\section*{IV. Radiation solution} 
To obtain solution of the simultaneous equations (12) and (13) under modified Nordtvedt conditions, we first write these equations in the synchronous gauge 
\begin{equation}
a^{\prime 2}+ka^{2}= \Gamma + \frac{A^{2}}{12  a^{2}}
\end{equation}
and 
\begin{equation}
2a a^{\prime \prime}-a^{\prime 2}+ka^{2}= -\Gamma + \frac{A^{2}}{4 a^{2}}
\end{equation}
where prime denotes $d/d\eta$ and the conformal time $\eta$ is defined through
\begin{equation}
ad\eta = dt
\end{equation}
For flat universe ($k=0$), solution of (16) and (17) is
\begin{equation}
a(\eta)= \left( \frac{A }{ \sqrt{3} } \eta + \Gamma \eta^{2}\right)^{1/2}\; ;
\end{equation}
and the cosmic time ($t$) is related with the conformal time ($\eta$) via 
\begin{equation}
t = \int{a d\eta}= \left(\frac{A}{4 \sqrt{3} \Gamma}+ \frac{\eta}{2} \right) \left( \frac{A}{ \sqrt{3}} \eta + \Gamma \eta^{2}\right)^{1/2} - \frac{A^{2}}{12\Gamma^{3/2}} Log \left [2 \sqrt{3 \Gamma \eta}+ 2\sqrt {3} \sqrt{ \frac{A}{ \sqrt{3}} + \Gamma \eta} \right]
\end{equation}
It is expected that all radiation solutions of ST theories with an arbitrary coupling function will converge to the above solutions under modified Nordtvedt conditions. For verification, we consider solutions obtained by Barrow for a class of ST theory [19] defined through 
$2\omega (\phi) +3 = 2B_{1}(1-\frac{\phi}{\phi_{0}})^{-\alpha}$; $\alpha>0$, $B_{1} >0$ constants. The theory reduces to Barker's ``constant'' $G$ theory [20] for $\alpha=1$ and $B_{1}=-1/2$ and to Brans-Dicke theory [6] for $\alpha=0$. The theory has been studied extensively in [3,5,19]. The exact radiation solutions for $\alpha=1$ is given by [19]
\begin{equation}
 \phi(\eta)=\frac{4\phi_{0}K^{\lambda}\eta^{\lambda}(\eta+2\eta_{0})^{\lambda}} {[(\eta+2\eta_{0})^{\lambda}+K^{\lambda}\eta^{\lambda}]^{2}}
\end{equation}
\begin{equation}
a^{2}(\eta)=\frac{\Gamma \eta (\eta+ \eta_{0}) }{\phi(\eta)}
\end{equation}
where $\lambda=(3/2B_{1})^{1/2}$, and $K$ and $\eta_{0}$ are constants. As  $\omega(\phi) \rightarrow \infty$, scalar field approaches $\phi_{0}$ and the expression for scale factor (22)  indeed converges to (19).  \\
It is clear from Eqs. (19) and (20) that for large $\eta$, $t \rightarrow \frac{\sqrt{\Gamma}}{2}\eta^{2}$ and $a \propto t^{1/2}$ i.e., the solution approaches the usual radiation dominated Friedmann model of GR. But when t is not very large, the scale factor differs from that of GR. 
\section*{V. Conclusion}
The evolution of scalar field suggests [3,4,5] that the coupling function $\omega(\phi)$ may evolve to a large value during radiation era. Some authors have discussed the effect of an inflationary phase in pushing $\omega(\phi)$ towards the extremum [3]. Consequently, there is a standard belief that ST theories are indistinguishable from GR at late radiation epoch and during matter dominated era. We have shown that the limit $\omega(\phi) \rightarrow \infty$ does not reduce a ST theory to GR for trace-free stress energy. But this is not surprising as the PPN parameters describe the possible deviations from GR in the local interaction of massive bodies and the Nordtvedt conditions reduce ST theories to GR only in the weak field limit. It is already known that the Brans-Dicke theory also does not always reduce to GR when $T=0$. But in that case the dependence of the BD scalar field $\phi$ on the coupling constant essentially remains arbitrary [16] and the asymptotic behavior of $\phi$ should be fixed respecting Machian nature of the BD theory (which is the motivation for the development of the theory) which immediately suggests $\phi \sim \phi_{0} + O \left( \frac {1}{\omega}\right)$ for large values of $\omega(\phi)$, so that the theory reduces to GR asymptotically [16,21]. But such a constraint on the functional dependence of $\phi$ on $\omega(\phi)$ is not allowed for general ST theories. \\
For compatibility with post-Newtonian weak field observations, Nordtvedt conditions are usually  imposed on ST theories. However, it is found that the original Nordtvedt conditions lead to an ambiguity in the value of Ricci scalar. As a result a slight modification of the second Nordtvedt condition is proposed (15) so that the Ricci scalar becomes finite and at the same time ST theories remain consistent with weak field observations. \\
The evolution of the scale factor during radiation era is studied for general ST cosmology under modified Nordtvedt conditions. An exact solution is obtained for flat universe which for certain conditions gives the corresponding GR solution. {\it All} classes (compatible with solar system observations) of ST radiation solutions should tend to this solution in the limit of large $\omega(\phi) $. At late times ($t \rightarrow \infty$) this solution tends to GR radiation solution but at early periods (when $t$ is not very large) it differs from the GR solution. The effect of such a difference on cosmological observations should be very interesting and will be investigated in a subsequent work. But it is evident that at least on principle the presence of a Brans-Dicke scalar field with an extremely large $\omega(\phi)$ could be detected through cosmological observations. \\

{\bf  \it Acknowledgment :}
The author wishes to thank IUCAA Reference Centre (NBU) for providing its facilities. It is a pleasure to thank G. Majumder (TIFR) for stimulating comments and various help.

\end{document}